\documentclass[epj]{svjour}
\usepackage{amssymb,amsmath,graphicx,color}

\newcommand{\ee}{\mathrm{e}}
\newcommand{\ie}{\emph{i.e.}}
\newcommand{\se}{_{\mathrm{e}}}		
\newcommand{\sn}{_{\mathrm{n}}}		
\newcommand{\sm}{_{\mathrm{m}}}		
\newcommand{\ol}[1]{\overline{#1}}
\newcommand{\avg}[1]{\langle #1\rangle}
\newcommand{\abs}[1]{\lvert #1\rvert}
\newcommand{\eref}[1]{Eq.~(\ref{#1})}
\begin{document}
\title{The effect of the initial network configuration on preferential attachment}
\author{Y. Berset and M. Medo\thanks{matus.medo@unifr.ch}}
\date{\today}
\institute{Department of Physics, Fribourg University, 1700 Fribourg, Switzerland}
\abstract{The classical preferential attachment model is sensitive to the choice of the initial configuration of the network. As the number of initial nodes and their degree grow, so does the time needed for an equilibrium degree distribution to be established. We study this phenomenon, provide estimates of the equilibration time, and characterize the degree distribution cutoff observed at finite times. When the initial network is dense and exceeds a certain small size, there is no equilibration and a suitable statistical test can always discern the produced degree distribution from the equilibrium one. As a by-product, the weighted Kolmogorov-Smirnov statistic is demonstrated to be more suitable for statistical analysis of power-law distributions with cutoff when the data is ample.
}
\maketitle

\section{Introduction}
The preferential attachment (PA) model proposed by Ba\-ra\-b\'asi and Albert is a network growth model where new nodes gradually appear and connect to existing nodes with probability proportional to the target node's degree~\cite{BarAlb99} (other frequently-used synonyms for this mechanism are rich-get-richer and cumulative advantage). Although not the first of its kind~\cite{mitzenmacher}, PA became popular for its simplicity and for producing a stationary power-law degree distribution which makes it a good candidate for modeling a wide range of real systems where heavy-tailed degree distributions are often observed~\cite[Ch.~3]{Newman2003}. The model helped to initiate the young field of complex networks~\cite{Newman2003,CohenHavlin10,Newman10} and it has been subsequently much studied and generalized (see in particular~\cite[Ch. 8]{AlbBar02} for an overview of analytical approaches to its solution and generalizations).

Significant evidence for preferential attachment has been found in various real datasets~\cite{Newman01,Jeong03,Leskovec08} but some important deviations have been reported too~\cite{AdaHub00,Redner05}, mainly in relation with the strong time bias of the model which causes that high degree nodes (the heavy tail) are almost exclusively those that were introduced in the early stage of the network's evolution. In the original PA model, if the network growth starts with two connected nodes (a so-called dyadic initial condition) and every new node creates one link, a node introduced at time step $i$ has at time $t$ expected degree $\sqrt{t/i}$ which decreases fast with $i$. (Since the distribution of nodes is uniform in $i$, this relation can be used to derive the well-known $1/k^3$ degree distribution in an especially simple way.) The drawback of time bias has been eliminated only recently by a model~\cite{MCG11} where aging of nodes makes it possible also for late introduced nodes to gain a significant number of links. Various other models of growing networks with aging of nodes exist and differ in their scope and behavior~\cite{DoMe2000,Eom2011}.

As networks rarely grow from a single starting node, we investigate the influence of an initial network of nodes on the original PA model. How is the stationary degree distribution formed and what is its functional form? To this end, we first show that if the degree of nodes in the initial network is greater than a certain threshold value (which we find to be approximately $3$), the initial nodes do not become part of the eventual power-law degree distribution of the network. To assess the approaching of the degree distribution of newly added nodes to a power-law form, we propose three quantities of interest and study their evolution with time. This leads to estimates of the distribution's equilibration time which are then interpreted in the context of the quantities used to obtain them.

When performing the goodness-of-fit of the network degree distributions, we find a divergence between results obtained with the Kolmogorov-Smirnov statistic used for statistical tests of power-law distributions in~\cite{Clauset2009} and those obtained with the weighted Kolmogorov-Smirnov statistic introduced in~\cite{Anderson1952}. We show that this difference is due to a cutoff of the network degree distributions and investigate the behavior and shape of this cutoff under various conditions. Our results reveal high sensitivity of the PA model to the initial network configuration which, to our best knowledge, has not been reported previously. Furthermore, significant differences exist between the ability of the standard and weighted Kolmogorov-Smirnov statistic to detect a power-law cutoff in empirical data. Note that finite size effects and sensitivity to the initial condition in the PA model have been studied already in~\cite{KraRed02} where however no results were provided for the equilibration time and the degree distribution cutoff.

\section{PA model with multiple initial nodes}
We study the PA model starting with an initial random network of $n_0$ nodes with mean degree $\mu_0$ where in every time step one node is added and creates a link to an existing node selected according to preferential attachment. The network thus consists of $n_0+t$ nodes after time step $t$. For the sake of clarity, nodes constituting the initial network are referred to as \emph{initial nodes} while all gradually added nodes are referred to as \emph{new nodes}.

The degree distribution of the initial nodes, $p_{k,t}$, can be studied by the standard master-equation approach~\cite{Levyraz}. Denoting the mean degree of the initial nodes at time $t$ as $\mu_t$, PA dictates that a link created at time step $t$ connects to one of the initial nodes with the probability $Q_t=(n_0\mu_t)/(n_0\mu_0+2t)$ where $n_0\mu_0+2t$ is the total degree of all nodes at time $t$. The master equation for $p_{k,t}$ follows in the form
\begin{equation}
\label{masteqn0}
p_{k,t+1} - p_{k,t} = \frac{(k-1) p_{k-1,t} - k p_{k,t}}{n_0\mu_0 + 2t}.
\end{equation}
By multiplying this with $k$ or $k^2$ and summing over all $k$, we obtain a difference equation for $\avg{k(t)}$ or $\avg{k(t)^2}$, respectively. A continuous time approximation then yields the average degree of the initial nodes, $\mu_t:=\avg{k(t)}$, and their average standard deviation, $\sigma_t:=\avg{k(t)^2}-\avg{k(t)}^2$, in the form
\begin{equation}
\label{initial_nodes-props}
\mu_t = \sqrt{\mu_0\left(\mu_0 + \frac{2 t}{n_0}\right)},\quad
\sigma_t = \sqrt{\mu_0 + \frac{2t}{n_0} - \mu_t}.
\end{equation}

\subsection{Separation of the initial nodes}
\begin{figure}
\centering
\includegraphics[width=\columnwidth]{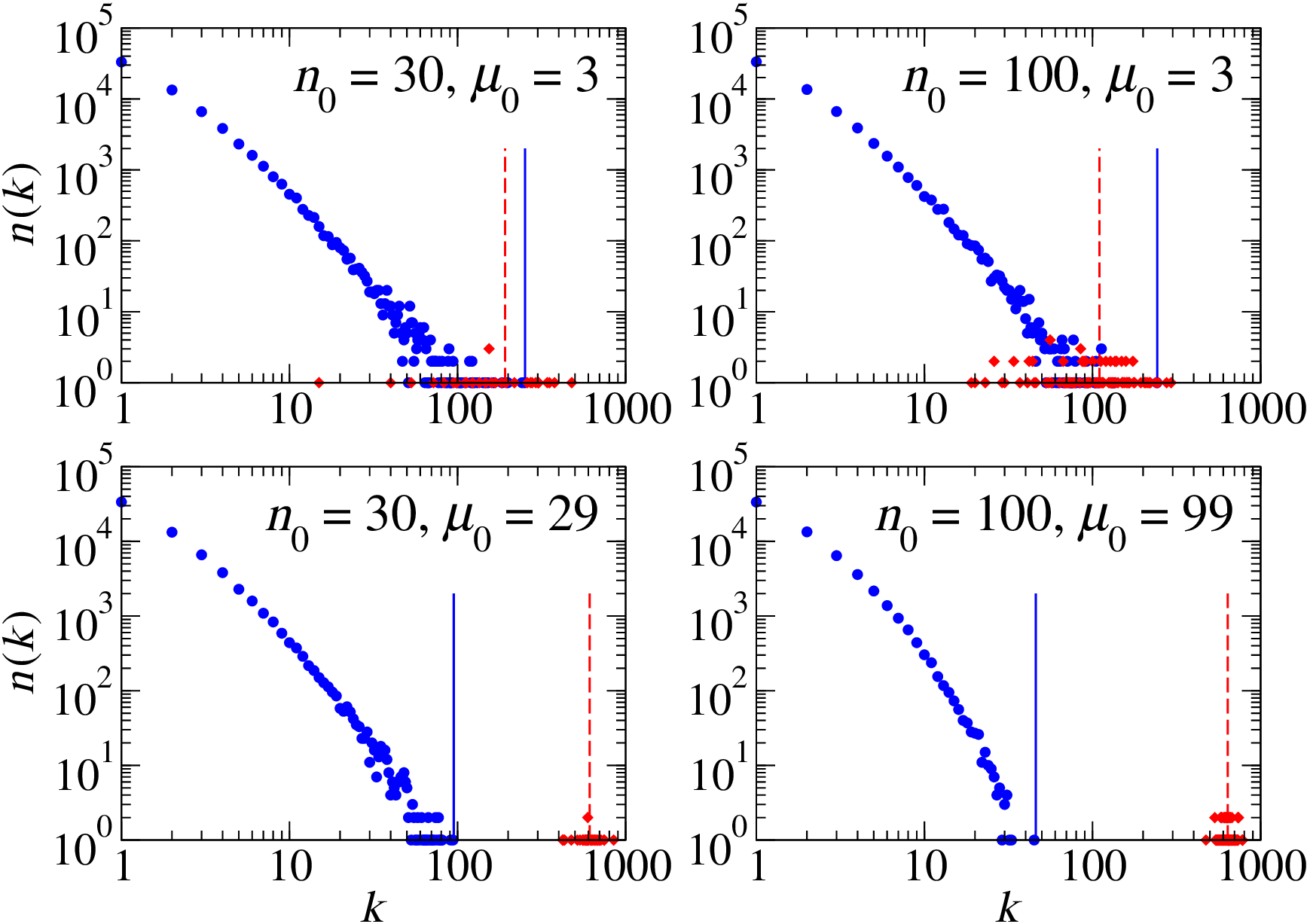}
\caption{Number of nodes of degree $k$, $n(k)$, for simulated PA networks with $t=2 \cdot 10^5$ added nodes: blue circles and red diamonds correspond to the new and initial nodes, respectively. Vertical lines mark the maximal degree of the new nodes (blue, solid) and the mean degree of the initial nodes (red, dashed). Separation of the initial nodes does not occur for $\mu_0=3$ (top) but it is clearly visible for two distinct choices of $n_0$ and $\mu_0$ where $\mu_0\gg3$ (down).}
\label{merging}
\end{figure}

We now examine whether the well-known stationary degree distribution of the original PA model
\begin{equation}
\label{BAdist}
f(k) = \frac{4}{k(k+1)(k+2)}
\end{equation}
can form in the presence of the initial nodes. To do that, we compare the number of the initial nodes with degree $\mu_t$ and the number of new nodes with this degree which, according to \eref{BAdist}, is $4t/[\mu_t(\mu_t+1)(\mu_t+2)]$. If the former number is greater than the latter, contribution of the initial nodes significantly distorts the expected form of $f(k)$ given above. Assuming that the degree distribution of the initial nodes is approximately Gaussian, there are roughly $n_0/\sqrt{2 \pi\sigma_t^2}$ of them with degree $\mu_t$. The initial nodes thus separate from the equilibrium degree distribution $f(k)$ when
\begin{equation}
\frac{n_0}{\sqrt{2 \pi} ~\sigma_t}\gtrsim \frac{4t}{\mu_t(\mu_t+1)(\mu_t+2)}
\end{equation}
Letting $t\to\infty$, we find that this inequality is always fulfilled for
\begin{equation}
\label{separation-condition}
\mu_0 \gtrsim 2\sqrt[3]{\pi} \approx 3.
\end{equation}
Hence regardless of the initial network size $n_0$ and the number of the new nodes $t$, the degree distribution of the initial nodes separates from that of the new nodes as long as $\mu_0\gtrsim 3$. Figure~\ref{merging} shows cases of merging and separation of the degree distributions for various values of $n_0$ and $\mu_0$. It confirms that when condition \eref{separation-condition} is met, the initial nodes remain well separated and visible in the degree distribution regardless of the values of $n_0$ and $t$. From now on, we thus focus on the degree distribution of the new nodes only and verify whether at least this can take the expected power-law form and when that happens.

\section{Equilibration time}
The degree distribution of the new nodes can be solved by master-equation in the large time limit. Despite the influence of the initial nodes at the beginning of the network's growth, the resulting distribution can be shown to be of the same form as for the original PA model, see \eref{BAdist}. To assess the time needed to achieve this equilibrium distribution, we employ three different approaches. For the sake of simplicity, we assume a complete initial network in this section, \emph{i.e.}, $\mu_0 = n_0-1$.

\subsection{Mean degree of the new nodes}
\label{sec:mean_deg}
In the early stage of the network's evolution, links from the new nodes initially frequently attach to the initial nodes. This causes the mean degree of the new nodes to be considerably lower than the overall mean degree which is two. Denoting the mean degree of the new nodes at time $t$ as $M_t$, the total number of links in the network, $n_0\mu_0+2t$ can be expressed as $tM_t + n_0\mu_t$. We can therefore use the previously obtained result for $\mu_t$ to obtain
\begin{equation}
M_t=2-\frac{n_0(\mu_t-\mu_0)}{t}
\end{equation}
which has the long time limit $M_{\infty} = 2$. To characterize the equilibration, we compute the time needed to reach $M_t=(1-\epsilon)M_{\infty}$ which follows in the form
\begin{equation}
t_{\mathrm{eq}} = \frac{(1 - 2\epsilon){n_0}^2}{2\epsilon^2} + O(n_0)\approx\frac{n_0^2}{2\epsilon^2}
\end{equation}
for large $n_0$ and small $\epsilon$. The equilibration time given by the mean degree of the new nodes thus grows quadratically with $n_0$. It is straightforward to verify that in the case of a general initial network with $\mu_0<n_0-1$, this result changes to $t_{\mathrm{eq}}\approx n_0\mu_0/(2\epsilon^2)$.

\subsection{Maximal degree of the new nodes}
\label{sec:max_deg}
We now consider the highest degree of a new node as an equilibration criterion. When the maximal degree observed in numerical simulations reaches the theoretically expected value following from the stationary distribution given by \eref{BAdist}, we say that the degree distribution has equilibrated.

\begin{figure}
\centering
\includegraphics[scale=0.28]{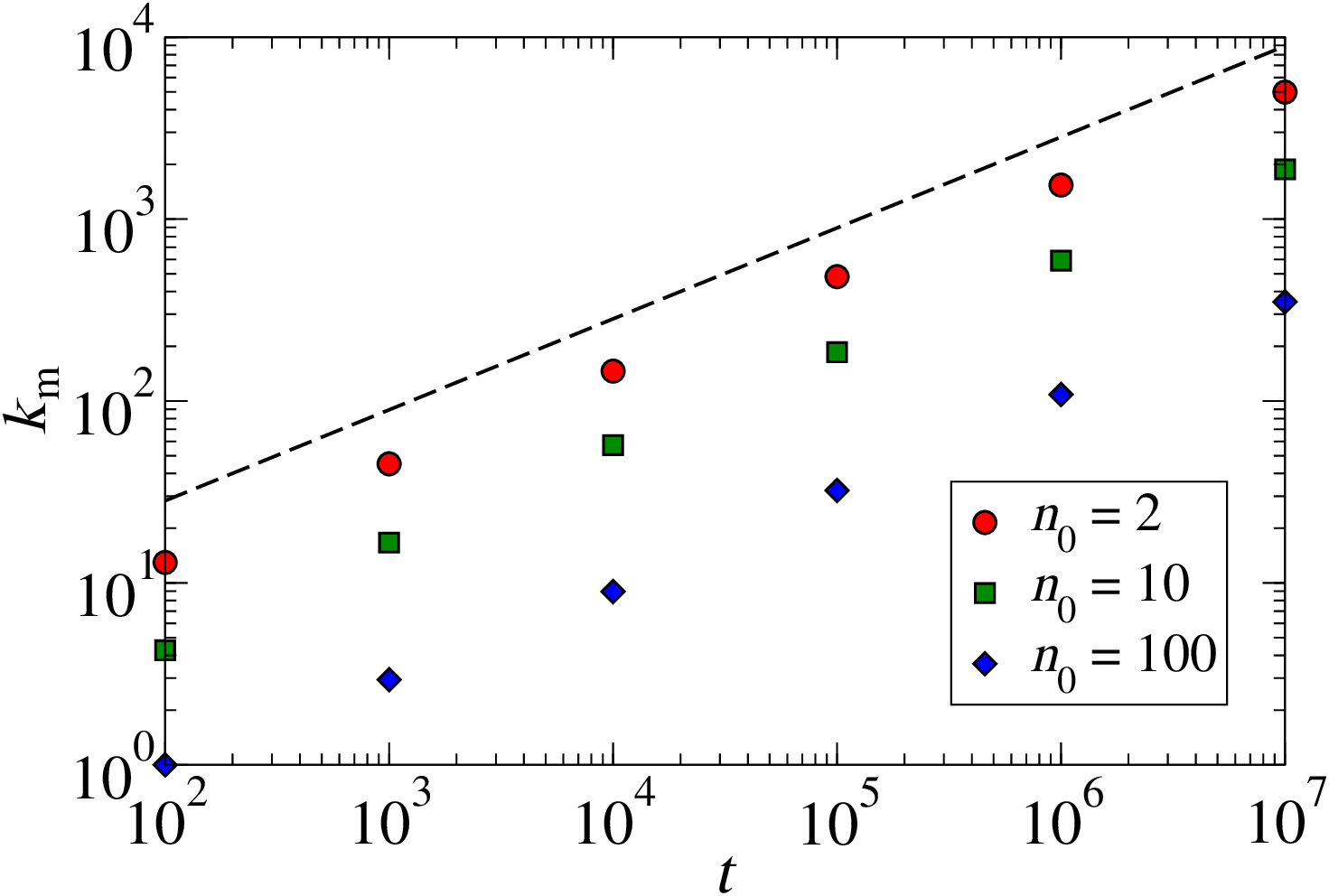}
\caption{Analytical results for the mean maximal degree (showed with the dashed line) and simulation results for the mean maximal degree at various values of $n_0$ (assuming a complete initial network, \ie, $\mu_0=n_0-1$). Results are averaged over 1000 network realizations.}
\label{fig:kmax}
\end{figure}

To compute the expected maximum degree value $\avg{k\sm}$, we study the extreme statistics for $t$ draws from the equilibrium distribution $f(k)$. Following the steps described in~\cite{Lu2009}, the probability that the highest degree value is $k\sm$ has the form
\begin{equation}
\label{kmax-start}
p(k\sm) = tf(k\sm) \left[\sum\limits_{k=1}^{k\sm-1} f(k)\right]^{t-1}.
\end{equation}
Approximating $(1-x)^{t-1}\approx \ee^{-x(t-1)}$ for small $x$, we get
\begin{equation}
\label{km-sum}
\avg{k\sm} = \sum\limits_{k\sm=1}^{\infty} \frac{4t}{(k\sm+1)(k\sm+2)}
\exp{\left[-\frac{2(t-1)}{k\sm(k\sm+1)}\right]}.
\end{equation}
This sum is easy to compute numerically but one can also estimate its value by roughly approximating the exponential term $\ee^{-ax}$ with one for $x\in[0,1/a]$ and zero for $x\in(1/a,\infty)$. This yields the expected value
\begin{equation}
\avg{k\sm} \approx \sqrt{8t}.
\end{equation}
A comparison of this result with a numerical summation of \eref{kmax-start} shows that when $t$ is large, the true value of $\avg{k\sm}$ is overestimated by less than $15\%$. While the average value of $k\sm$ following from simulations, $\ol{k\sm}$, is also proportional to $\sqrt{t}$, it always holds that $\avg{k\sm}>\ol{k\sm}$ and the gap between the two quantities grows with the number of the initial nodes $n_0$ (see Figure~\ref{fig:kmax}). We can conclude that no equilibration time can be defined here and the extreme degree statistics suggests that the degree distribution of the new nodes never reaches the stationary form prescribed by \eref{BAdist}.

\subsection{Fitting the network degree distributions}
We finally study the agreement between functional forms of the simulated and the equilibrium degree distribution, respectively. The standard approach to this task is a so-called goodness-of-fit test. Given a set of observed data and an expected statistical distribution, it measures how much the data fluctuates from the expected distribution compared to artificial data drawn from this distribution. In particular, we adopt a procedure presented in~\cite{Clauset2009} especially for statistical analysis of power-law distributions which goes as follows. For an input realization of the network at time $t$ (\emph{i.e.}, after adding $t$ new nodes), one computes the cumulative degree distribution of the new nodes, $R(k)$, and the cumulative degree distribution of the expected distribution, $T(k):=\sum_{k'=k}^{\infty} f(k')$. The Kolmo\-go\-rov-Smirnov statistic (KS) introduces the distance between the two cumulative distributions
\begin{equation}
D_0 = \max\limits_k\,\abs{T(k)-R(k)}.
\end{equation}
One then generates a large number of artificial datasets following the expected distribution and having the same size as the input data and computes the Kolmogorov-Smirnov statistic $D_1$ for them. The fraction of datasets with $D_1>D_0$ then gives $p$-value of the fit between the input degree distribution $R(k)$ and the expected degree distribution. By averaging this result over various realizations of the network, we obtain the final $p$-values which are reported here. We significantly speed up the computation by using the same set of artificial datasets and their $D_1$ values to evaluate each network realization at a given time $t$. The hypothesis that the network degree data follows the expected distribution \eref{BAdist} is then evaluated on the basis of the resulting $p$-value. If $p<0.1$, the hypothesis is rejected. In other words, the hypothesis of agreement is plausible as long as at least 10\% of the the artificial data agree less with the expected distribution than simulated network data do. The same procedure can be carried out using the Anderson-Darling statistic~\cite{Anderson1952} (which is referred to as weighted Kolmogorov-Smirnov statistic (WKS) in~\cite{Clauset2009})
\begin{equation}
D^* = \max\limits_{k} \frac{\abs{T(k)-R(k)}}{\sqrt{T(k)[1-T(k)]}}.
\end{equation}
The corresponding $p$-value is denoted $p^*$.

\begin{figure}
\centering
\includegraphics[scale=0.28]{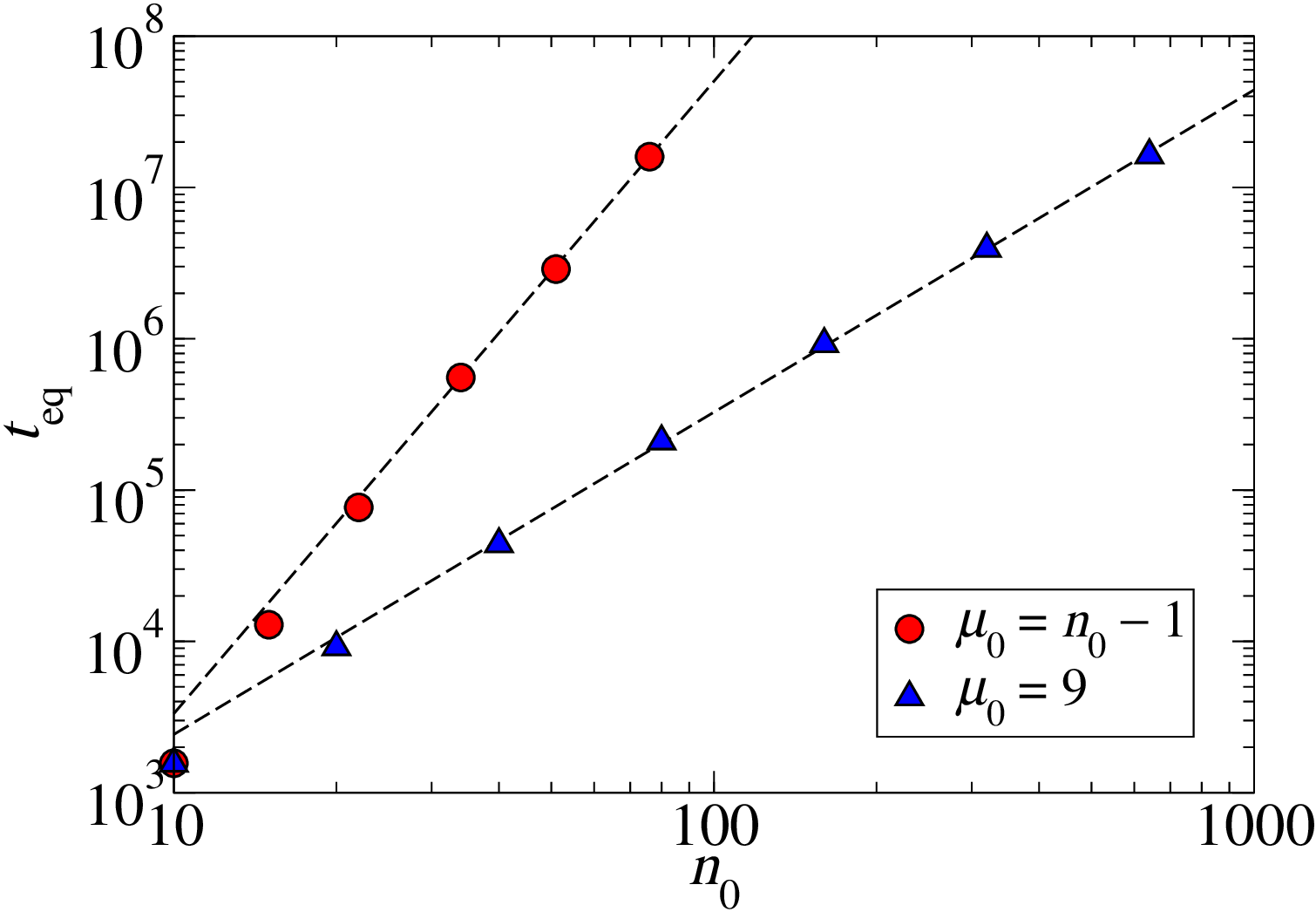}
\caption{Scaling of the $p$-value-based equilibration time $t_{\mathrm{eq}}$ with the number of initial nodes $n_0$ for the complete initial network ($\mu_0=n_0-1$) and the initial network with fixed degree ($\mu_0=9$). Numerical results and corresponding linear fits are shown with symbols and dashed lines, respectively. The dashed lines have slopes $4.18$ and $2.13$, respectively.}
\label{fig:eq_time}
\end{figure}

Equilibration time can be defined based on when $p$ reaches the threshold value of $0.1$ and the stationary distribution $f(k)$ therefore becomes a plausible hypothesis for simulated networks. Figure~\ref{fig:eq_time} shows that $t_{\mathrm{eq}}$ scales with $n_0$ as $t_{\mathrm{eq}}\sim n_0^{\beta}$ where $\beta=4.18\pm0.02$ for $\mu_0=n_0-1$ (applies for $n_0\geq 30$) and $\beta=2.13\pm0.01$ for constant $\mu_0$ (applies for $n_0\geq 40$). Note that similarly as before, we have the scaling exponent for complete initial networks twice as high as for initial networks with fixed $\mu_0$. What is different from equilibration based on the average degree of the new nodes is that for both fixed and growing $\mu_0$, we observe much faster growth of $t_{\mathrm{eq}}$ with $n_0$.

Very recently, a new goodness-of-fit test has been proposed~\cite{Bouchaud12} which also relies on the KS statistic but circumvents the $p$-value testing. This approach is distribution-free and focuses only on whether the KS statistic of a data set is higher than a certain threshold value. In particular, the hypothesis that the given data follows a power-law distribution can be discarded with 90\% confidence when its KS statistic $D_0$ is higher than $1.224/\sqrt{t}$ for data size $t$ (for 95\% confidence level, the threshold would be $1.358/\sqrt{t}$ as reported in~\cite{Bouchaud12}). Besides saving computational time (no artificial data sets need to be generated here), this method provides scaling exponents $\beta$ that match well with the ones derived above. We can conclude that a very long time is needed to achieve network degree distributions that are accepted to be compatible with the equilibrium degree distribution by the standard Kolmogorov-Smirnov test.

\begin{figure}
\centering
\includegraphics[scale=0.28]{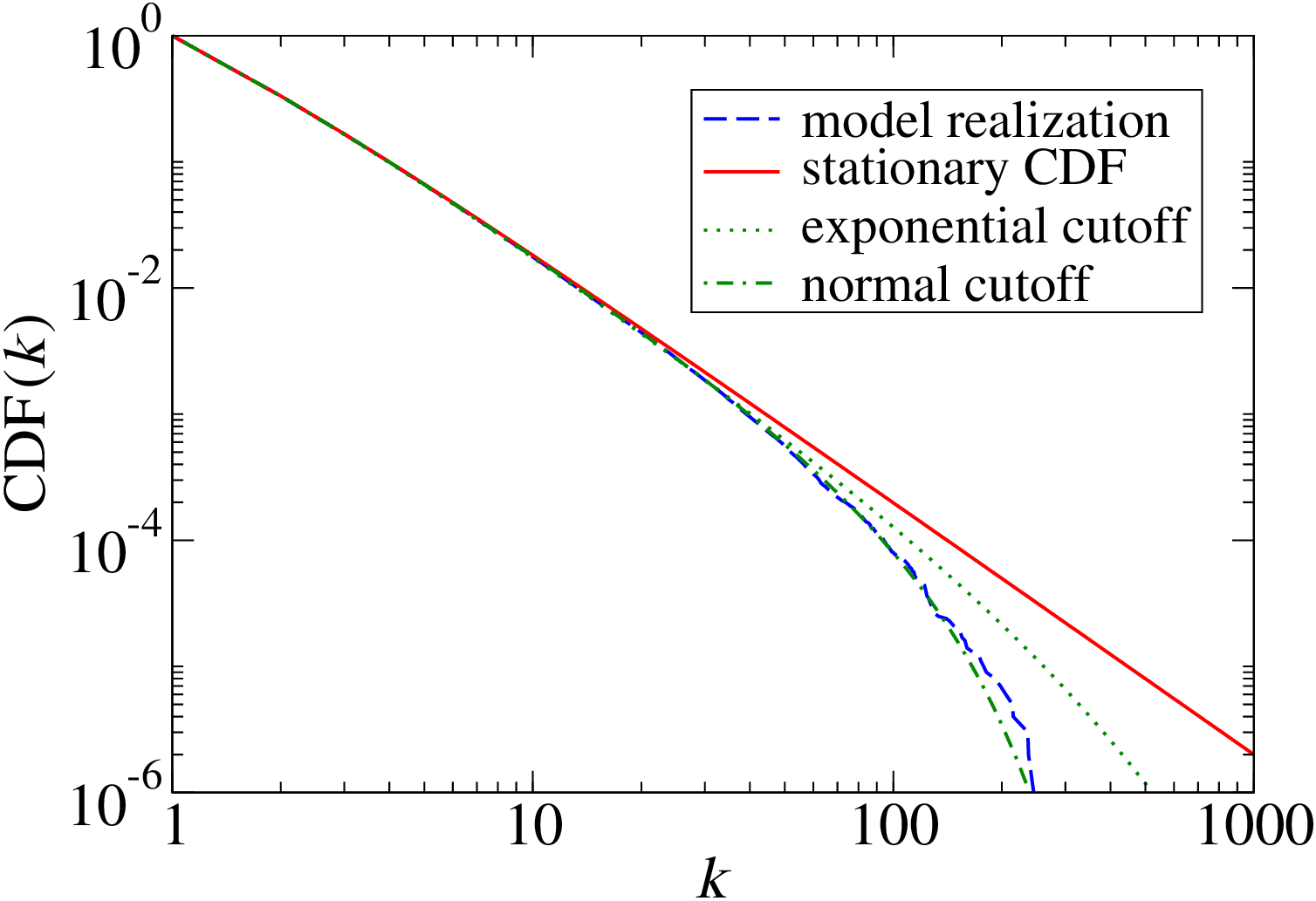}
\caption{The cumulative degree distribution of one network realization for $n_0=30$, $\mu_0=29$, and $t=10^6$ (dashed line) and the stationary distribution (solid line). In this case, the two variants of the goodness-of-fit test provide contradictory values $p=0.30$ and $p^*=0.01$. Fits of the degree distribution with exponential and normal cutoff (see Section~\ref{sec:cutoff}) are also shown here (the corresponding $p$ and $p^*$ values, averaged over multiple network realizations, can be found in Figure~\ref{fig:pvalues}).}
\label{fig:CDFs}
\end{figure}

\begin{figure*}
\begin{center}
\includegraphics[scale=0.56]{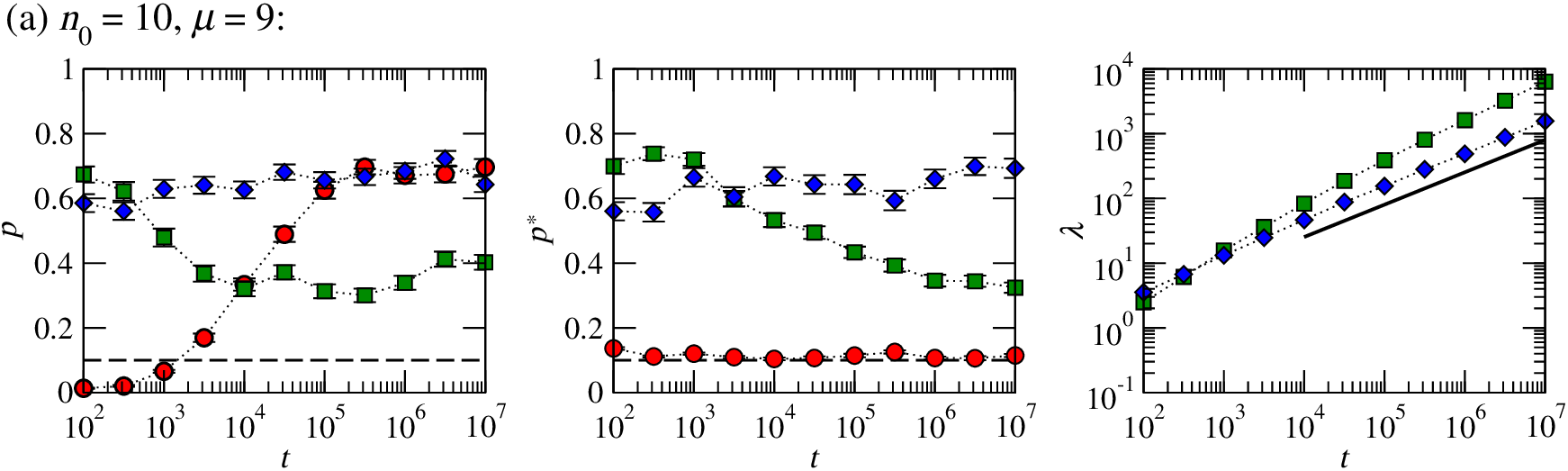}\\[6pt]
\includegraphics[scale=0.56]{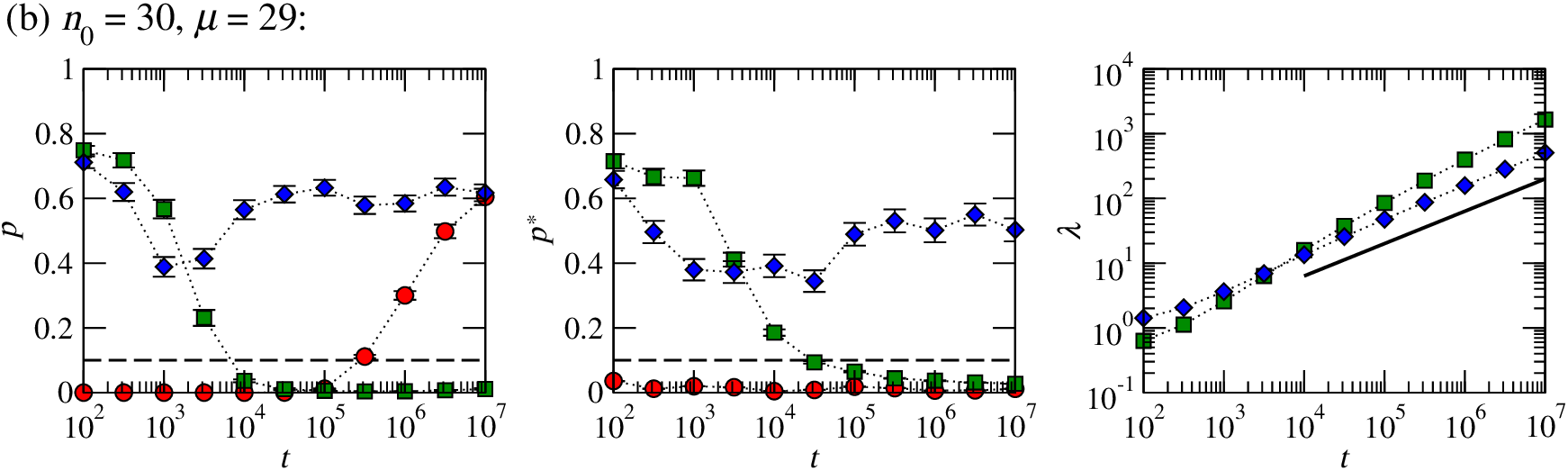}\\[6pt]
\includegraphics[scale=0.56]{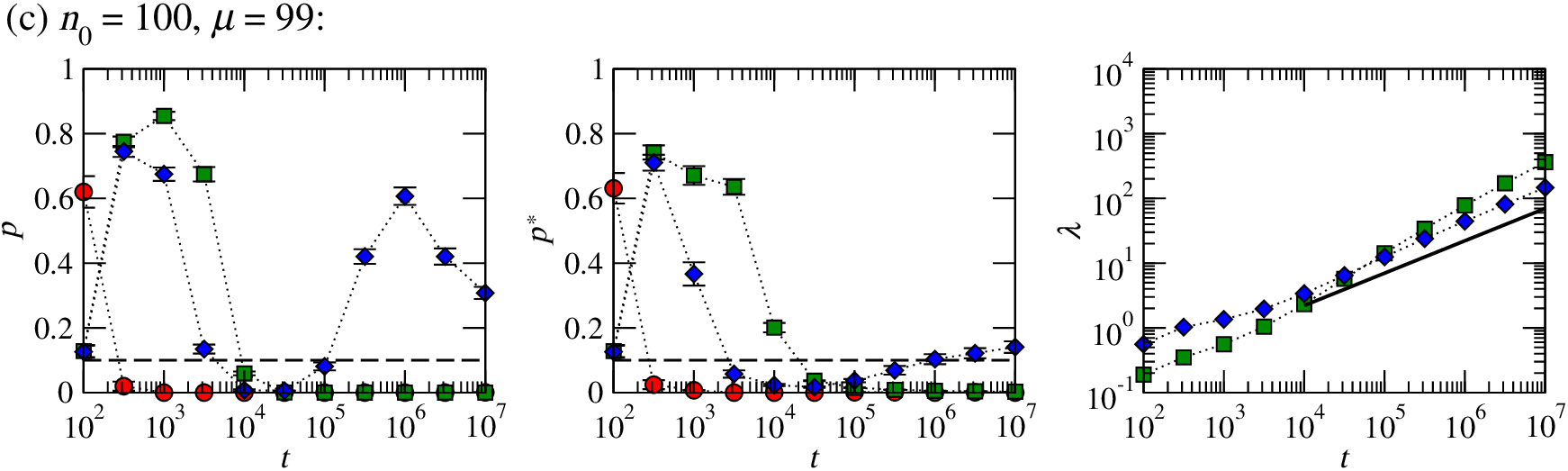}\\
\caption{$p$-values, $p^*$-values, and cutoff values $\lambda$ versus $t$ for $n_0=10$, $\mu_0=9$ (top row), $n_0=30$, $\mu_0=29$ (middle row), and $n_0=100$, $\mu_0=99$ (bottom row) with different symbols corresponding to the fitting of different degree distributions: stationary degree distribution of the PA model $f(k)$ (red circles), $f(k)$ with exponential cutoff (green squares), and $f(k)$ with normal cutoff (blue diamonds). Horizontal dashed lines mark the threshold $p$-value of $0.1$. Results are averaged over 100 network realizations, each of which is compared with 1000 draws from the reference distribution. Thick solid lines in the graphs of $\lambda$ serve as guides to the eye and have all slope $0.5$.}
\label{fig:pvalues}
\end{center}
\end{figure*}

While $p$-values follow the expected scenario and grow with $t$, thus allowing a new equilibration time to be introduced, simulations show that $p^*$-values based on the WKS are essentially independent of $t$. As soon as $\mu_0\gtrsim10$, $p^*<0.1$ for any value of $t$ (except for very low $t$ where however high values of $p^*$ are due to fluctuations of the tiny evaluated data)---see the corresponding lines in Figure~\ref{fig:pvalues}. To understand what causes this behavior, it is instructive to plot the cumulative network degree distribution and compare it with the stationary distribution. This is shown in Figure~\ref{fig:CDFs} where one can see that the tails of these two distributions differ substantially with the network degree distribution showing cutoff for degree greater than approximately $30$. Note that this cutoff is exactly the reason why the observed $k\sm$ values reported in Figure~\ref{fig:kmax} are lower than expected. Despite the difference in CDFs, the goodness-of-fit leads to a threshold-satisfying $p$-value $0.25$ which suggests a high degree of agreement according to the KS statistic. This inability of the KS to detect the deviation between the distributions is because it is based only on the differences between CDFs which are inevitably small at the tail (distance $\abs{T(k)-R(k)}$ cannot exceed $\max\{T(k),R(k)\}$). By contrast, the WKS is weighted by $1/\sqrt{T(k)[1-T(k)]}$ which makes it more sensitive to CDF differences that occur in the tail where $T(k)$ is small and allows it to reject the hypothesis of the network degree distribution being compatible with the stationary distribution with $p^*\approx0.01$. Note that the approach proposed in~\cite{Bouchaud12} can be applied also to the WKS and again agrees with the findings presented here.

\subsection{Cutoff fitting}
\label{sec:cutoff}
Given the sensitivity of the WKS to the tail behavior, it is now natural to use it to study the cutoff type and position as a function of $n_0$ and $t$. In addition to the usual exponential cutoff which is often seen in real data~\cite{Clauset2009}, we test also a so-called normal cutoff of the form $\exp[-(k/\lambda)^2]$ which is a special case of the stretched exponential function (sometimes it is referred to as compressed exponential function because it decays faster than exponentially). This choice is further supported by likelihood of the network degree data: when the cutoff term is assumed in the form $\exp[-(k/\lambda)^{\gamma}]$, likelihood of the data reaches its maximum for $\gamma$ between $1.5$ and $2.5$ (as $t$ grows, the maximum shifts to higher values). We thus have two candidate distributions
\begin{align}
\label{cutoff_distrib}
f\se(k)&=\frac{A(\lambda\se)\,\ee^{-k/\lambda\se}}{k(k+1)(k+2)},\\
\label{cutoff_distrib2}
f\sn(k)&=\frac{B(\lambda\sn)\,\ee^{-(k/\lambda\sn)^2}}{k(k+1)(k+2)}
\end{align}
where $A(\lambda\se)$ and $B(\lambda\sn)$ are normalization factors. The procedure is now as follows. For a particular network realization, one chooses the cutoff parameter that maximizes likelihood of the data (taking only the new nodes into account). $p$- and $p^*$-value are then computed with respect to Eqs.~(\ref{cutoff_distrib}) and (\ref{cutoff_distrib2}) as reference distributions. By averaging over various network realizations, we obtain statistics for $\lambda\se$ and $\lambda\sn$ as well as average values of $p$ and $p^*$ which measure the goodness of fit.

Figure~\ref{fig:pvalues} summarizes results of the cutoff analysis. First of all, it shows the previously mentioned fact that while $p$-values obtained for the stationary cutoff-free distribution increase fast with $t$, $p^*$-values of this fit are low and insensitive to $t$. Fits with exponential cutoff perform better than the original stationary distribution with respect to both $p$ and $p^*$ but both quantities gradually decrease with $t$ instead of increasing (which is an unexpected behavior because the fit is supposed to improve as the network grows). Finally, the normal cutoff performs best and its $p$ and $p^*$ values do not decay with $t$. One may wonder how is it possible that distributions with cutoff are able to achieve high $p$ and $p^*$ values even when $t$ is very small and the core part of the degree distribution, $1/(k(k+1)(k+2))$, has not yet had the time to form. The reason lies in very small cutoff values inferred by likelihood estimation in those cases (see the values shown in panels in the last column in Figure~\ref{fig:pvalues}) which results in the distribution shape being dominated by the cutoff part instead of the previously-mentioned core part. We finally note that for normal cutoff, the cutoff parameter values are proportional to $t^{0.5}$. This is the same scaling as we found earlier for $\avg{k\sm}$. This is understandable: normal cutoff is sharp and its position is mainly influenced by the highest degree values occurring in the network.

\subsection{Comparison with an analytical solution}
After the original submission of our manuscript, an analytical work has been published where Z-transform is used to find the degree distribution as a function of time for a growing network with an arbitrary initial condition~\cite{FoRa12}. When the network growth obeys preferential attachment, their final result given in Eqs.~(81) and (82) can be adapted to our setting and yields the degree distribution of the new nodes in the form
\begin{equation}
\label{Pk-analytical}
P(k,t) = \frac{n_0\mu_0+2t}{t}\left(\frac1k-\frac{2c}{k+1}+\frac{c^2}{k+2}\right)c^k
\end{equation}
where $c=1-\sqrt{n_0\mu_0/(n_0\mu_0+2t)}$. (The first term of Eq.~(81) does not appear here because it describes the contribution of the initial nodes. The normalization is changed from $1/(\mu_0+t)$ to $1/t$ because our $P(k,t)$ covers $t$ new nodes instead of all $\mu_0+t$ nodes as in~\cite{FoRa12}.) This result agrees well with our simulations.

When $t\to\infty$, $c=1$ and $P(k,t)$ reduces to \eref{BAdist} as it has to. However, one can write $c=1-\sqrt{x^2/(1+x^2)}$ where $x:=\sqrt{n_0\mu_0/(2t)}$ and consequently find an expansion of $P(k,t)$ in powers of $x$. The leading order part of the result,
\begin{equation}
\label{expansion}
P(k,t)=\frac{4\left(1-\tfrac16 (kx)^3+\tfrac18 (kx)^4 + O\big((kx)^5\big)\right)}{k(k+1)(k+2)},
\end{equation}
contains the stationary solution and correction terms proportional to $kx$ and its powers. While $x$ vanishes as $t\to\infty$, the growing network allows us to inspect $P(k,t)$ at higher values of $k$. Assuming that the stationary distribution eventually establishes itself over the whole range of relevant degrees, the expected largest degree is $\avg{k\sm}\approx\sqrt{8t}$ (as shown in Section~\ref{sec:max_deg}). This means that the correction terms $k\sm x$ are independent of $t$ and thus do not vanish: a deviation between the stationary distribution and the ``visible part'' of $P(k,t)$ persists. The analytical form of $P(k,t)$ given in \eref{Pk-analytical} thus confirms the statistical tests of model degree distributions reported above.

\section{Conclusion}
The lack of attention to the importance of initial conditions in network models is best illustrated by thirteen years separating the original publication of the preferential attachment model~\cite{BarAlb99} and the analytical result for the model's degree distribution upon arbitrary analytical conditions~\cite{FoRa12}. We studied the sensitivity of the Barab\'asi-Albert model of a growing network to the initial network from which the growth starts. We found that the well-known stationary distribution $f(k)=4/[k(k+1)(k+2)]$ forms only when the number of the initial nodes are few and they are sparsely interconnected. We demonstrated that as soon as the starting degree of the initial nodes $\mu_0$ exceeds $3$, this little advantage allows them to attract an excessive number of links in the future so that they never merge with the stationary degree distribution of the nodes that are introduced later in the network's evolution.

When focusing only on the newly added nodes and their degree, we showed that their stationary degree distribution is the same as that of the original model regardless of the number of initial nodes $n_0$ and their degree $\mu_0$. There are various ways how to define the time needed to approach this distribution. If we define the equilibration time simply on the basis of the average degree of the new nodes, it is proportional to $n_0^2$ in the case of the complete initial network (and proportional to $n_0\mu_0$ in general) which suggests rather fast equilibration. On the basis of the standard goodness-of-fit test with the Kolmogorov-Smirnov statistic, the equilibration time grows with $n_0$ much faster---the exponent is around $4.2$ for the complete initial network and $2.1$ when $\mu_0$ is fixed.

However, no equilibration is found in two other cases which are in fact closely related. In the first case, we showed that when $n_0\gtrsim10$, the average maximal degree of the new nodes is and stays significantly smaller than the value predicted from the stationary distribution. In the second case, we showed that when the usual Kolmogorov-Smirnov statistic is replaced by the weighted Kolmogorov-Smirnov statistic which puts more weight on the tail of a distribution, the hypothesis that the network degree distributions are drawn from the stationary distribution of the PA model is rejected for $n_0\gtrsim10$ (for complete initial networks). The reason for these two observations lies in a distribution cutoff which shifts to higher degree values as the network grows (thus the eventual convergence to $f(k)$ in the functional form) but remains present and detectable for any finite network size. One can conclude that with respect to more sophisticated equilibration criteria, degree distributions of the PA model equilibrate slowly (with respect to the KS) or they do not equilibrate at all (with respect to the WKS). These results are confirmed by a recently published analytical form of the degree distribution of the PA model for an arbitrary initial condition. Note that models of network growth where aging of nodes is considered~\cite{MCG11,DoMe2000,Eom2011} naturally depend less on the initial network configuration. One can thus expect that these models not only solve the problem of node degree strongly biased by time (as is the case for PA) but also that of the lack of equilibration. We studied also other common network characteristics, clustering coefficient and assortativity, and found that their overall behavior is not altered by the presence of a non-trivial initial network. They both vanish in the limit of $t\to\infty$, albeit at rates which depend on $n_0$ and $\mu_0$.

\begin{figure}
\begin{center}
\includegraphics[scale=0.28]{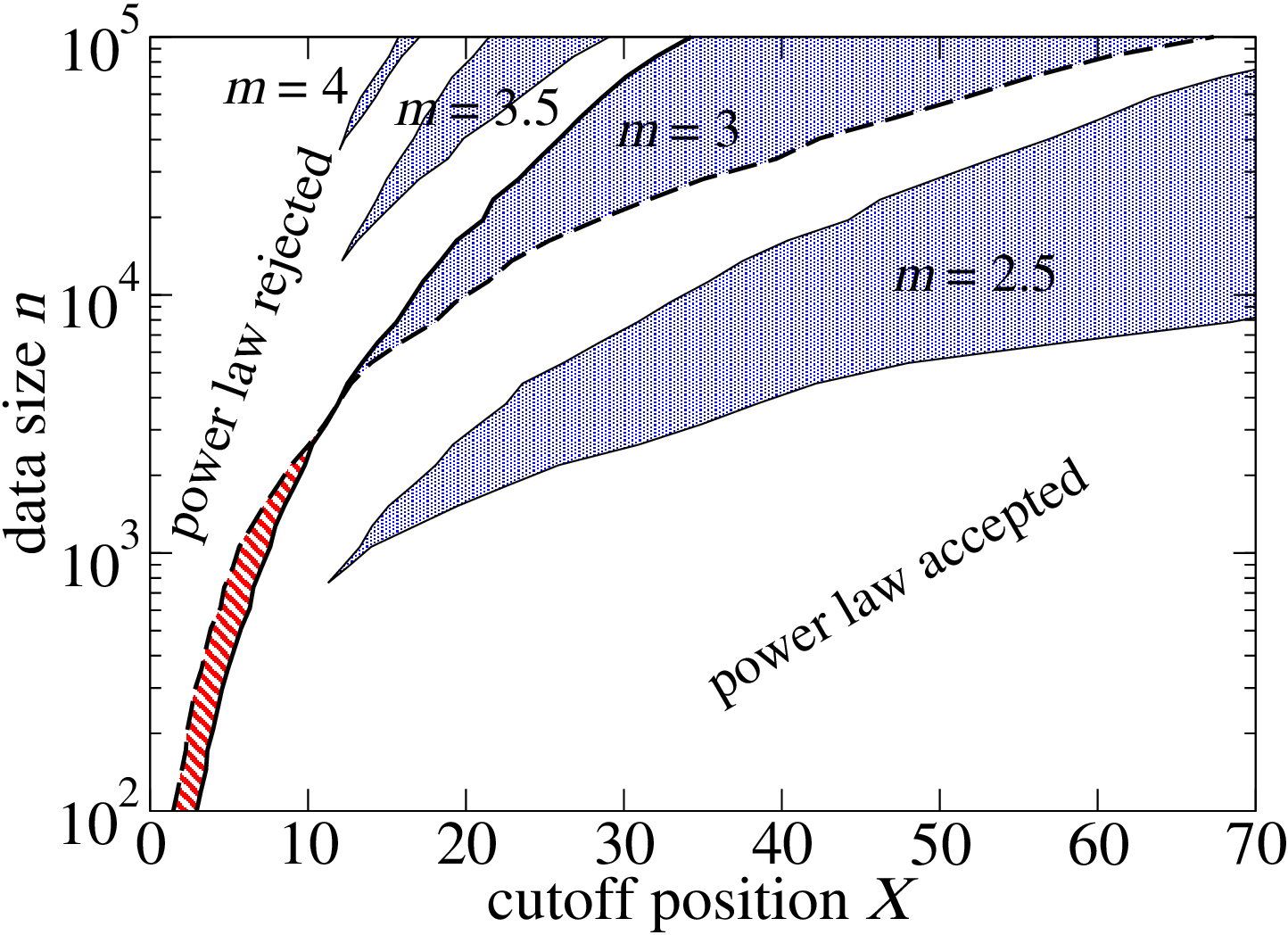}
\caption{Data size $n$ needed for the assumed power law $k^{-m}$ to yield $p$ or $p^*$ less than $0.1$ (the hypothesis is rejected) when the input data follows $k^{-m}\exp[-(k/X)^2]$. Results obtained for $m=3$ with $p$ and $p^*$ are shown with the thick solid and dashed line, respectively. $p^*$ outperforms $p$ in blue-shaded regions (shown for $m=2.5,3,3.5,4$). For small $n$ and $X$, there are also regions where $p$ outperforms $p^*$ (for clarity shown only for $m=3$ and marked with red stripes).}
\label{fig:phase_diagram}
\end{center}
\end{figure}

We finally stress that there is a more general lesson to be learned here. Despite the conventional wisdom~\cite{Clauset2009}, the standard and weighted Kolmogorov-Smirnov statistic may perform very differently on power-law data with cutoff. When the cutoff is located at large values of a variable, it may remain invisible to the standard Kolmogorov-Smirnov statistic which then accepts data as being plausibly generated by a given distribution. By contrast, sensitivity of the WKS statistic is distributed more evenly over the range of possible values which improves its ability to detect cutoffs and estimate their parameters (such as position and shape, for example). This is demonstrated in Figures~\ref{fig:phase_diagram} where the data size needed to reject the power law hypothesis for a data generated by a power law with normal cutoff is shown as a function of the cutoff position. When the data is big enough, the $p^*$-value test can ``detect'' higher cutoff values than the $p$-value test which makes it a preferable choice in a wide range of parameters. Regions where $p^*$ outperform $p$ are smaller when the actual cutoff has an exponential form. When the data is small ($n\lesssim 1000$) or the power-law exponent is high (four or more), it is still advisable to use the standard Kolmogorov-Smirnov statistic.

\begin{acknowledgement}
This work was supported by the EU FET-Open Grant No. 231200 (project QLectives) and by the Swiss National Science Foundation Grant No. 200020-132253 (project Evolving and adaptive networks).
\end{acknowledgement}

\end{document}